*I.A.Vereshchagina*
*(Pulkovo Observatory, St.-Petersburg, Russsia)*

*The results of studies of 2006 VV2 and 13170 (1999 HF1) binary asteroids of NEA group, (45) Eugenia and (87) Sylvia triple asteroids, (762) Pulcova and (90) Antiope binary asteroids of the Main Belt are presented. For the asteroid 2006 VV2, color indices (B–V, V–R, R–I), the absolute magnitude (16.7 ± 0.2$^m$), a taxonomy class, the density (2.71 ± 0.04 g/cm$^3$), the masses of components as well as refined form of the main component and the satellite orbit were determined. For the (45) Eugenia asteroid, refined form the main component, simulated orbits of both satellites were determined and it was showed that the axis of rotation of the main component undergoes a forced precession with an angle of 10° and a period of 66 days. Based on a comparison of model lightcurves with observations over a long interval of time of (90) Antiope asteroid, reflective surface properties of its components were investigated. For asteroid (762) Pulcova, form of the main component was determined and the reflective properties of its surface were investigated. For asteroid 137170 (1999 HF1), estimates of the size of the main component (a = 2.12 ± 0.02, b = 1.77 ± 0.01, c = 1.73 ± 0.03 km) were made and possible stable orbit of the satellite (with the elements of a ~ 6.2 km, e ~ 0.1, i ~ 2°) was identified and of component masses were estimated. For the asteroid (87) Sylvia, form the main component was refined.*

## Methods of the investigation

To study selected asteroids, photometric observations of these objects on the time interval from 2006 to 2010 were performed. The observations were made with ZA-320M and MTM-500M automated telescopes of Pulkovo Observatory [Devyatkin et al, 2009a]. Processing of the observations was carried out with Apex-I and Apex-II software [Devyatkin et al, 2009b].

To simulate multiple asteroids, equations of translational and rotational motion for the two and three bodies respectively were used. In modeling the light curve, we have used the Lumme-Bowell [Bowell et al, 1989; Karttunen, 1989] and Hapke [Hapke, 1981a, 1981b, 1981c, 1986, 2002] laws of reflection. To restore the form of an asteroid using the observed light curves we have used a method proposed by Kaasalainen [Kaasalainen et al, 2001a, 2001b].

## Binary asteroid 2006 VV2 group of NEA

For this asteroid on the basis of these observations the color indices (Table 1), the absolute magnitude (Table 2) and its possible taxonometric class A were determined. The color indices were determined by two methods described in [Vereshchagina et al, 2009]. Based on these results, it was estimated the density of the system which amounted to 2.71 ± 0.04 g/cm$^3$. Using this value for the first time estimates of the components masses were obtained. They are listed in Table 3.

**Table 1.** Estimates of the color indices of asteroid 2006 VV2.

|  | *B–V* | *V–R* | *R–I* |
|---|---|---|---|
| **1-st method** | 0.67 ± 0.14 | 0.45 ± 0.09 | 0.18 ± 0.11 |
| **2-nd method** | 0.88 ± 0.07 | 0.50 ± 0.09 | 0.13 ± 0.10 |
| **Betzler et al, 2008** | - | 0.64 ± 0.07 | - |
| **Betzler et al, 2009** | 0.84 ± 0.06 | 0.39 ± 0.02 | - |
| **Hergenrother et al, 2009** | 0.79 ± 0.01 | 0.45 ± 0.01 | 0.37 ± 0.01 |

**Table 2.** Estimates of the absolute magnitude of asteroid 2006 VV2.

|  | $H_V$ |
|---|---|
| **Our result** | 16.7 ± 0.2 |
| **JPL** | 16.8 ± 0.5 |
| **Betzler et al, 2009** | 16.6 ± 0.2 |

Based on the observational data evaluation of the main component shape of the asteroid was determined, the position of the pole of rotation was defined and the period of axial rotation was specified. The corresponding results are shown in Table 4. Figure 1 shows the obtained shape of the main component in three different viewpoints.

**Table 3.** Estimates of the components masses of the 2006 VV2 asteroid in assumption that the density is 2.71 ± 0.04 g / cm$^3$ (classes A, Q, V).

|  | **Mass, kg** |
|---|---|
| **Main component** | 8.275 ×10$^{12}$ ± 0.122 ×10$^{12}$ |
| **Satellite** | 1.77 ×10$^{11}$ ± 0.03 ×10$^{11}$ |
| **System mass** | 8.45 ×10$^{12}$ ± 0.13 ×10$^{12}$ |

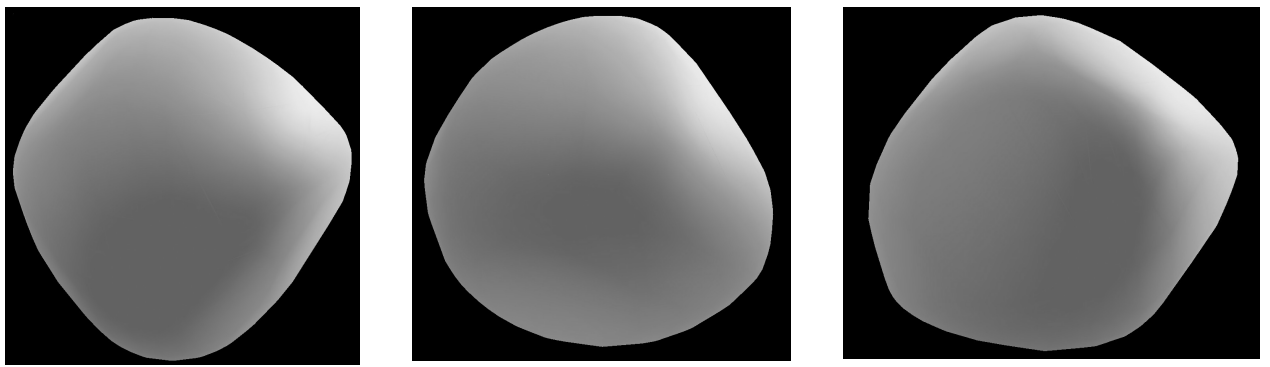

**Fig. 1.** The shape of the main component of the 2006 VV2 asteroid shown in the three different viewpoints.

Obtained shape of the main component and estimates of the masses allowed to determine the possible stable orbit of the satellite. It is the closest orbit to the data of radar observations. Table 5 shows the elements of the orbit.

**Table 4.** Estimates of the ecliptic coordinates of the pole, the rotation period and the dimensions of the main component of the 2006 VV2 asteroid.

| β, ° | λ, ° | *P*, h |
|---|---|---|
| 37 ± 2 | 29 ± 3 | 2.410541 ± 0.000003 |

| Dimensions $a \times b \times c$, km | $0.92 \times 0.89 \times 0.89 \ \pm\ 0.05$ |
|---|---|

**Table 5.** Elements of stable orbit of 2006 VV2 asteroid. Also estimates of orbital parameters taken from IAU Circular 8826, 2007 are presented.

|  | Semimajor axis $a$, km | Eccentricity, $e$ | Inclination, $i$, ° | Period $P$, h |
|---|---|---|---|---|
| **IAU Circular 8826, 2007** | $\geq 1.5$ | - | - | $\sim 5$ |
| **Our results** | $1.9 \pm 0.2$ | $0.10 \pm 0.06$ | $0.0 \pm 0.002$ | $6.1 \pm 0.2$ |

The constructed model of 2006 VV2 asteroid allow to obtain a model lightcurve of the system. The comparison of this model with the observed lightcurves is shown in Figure 2. One can see that the lightcurves (model and observed) are in good agreement.

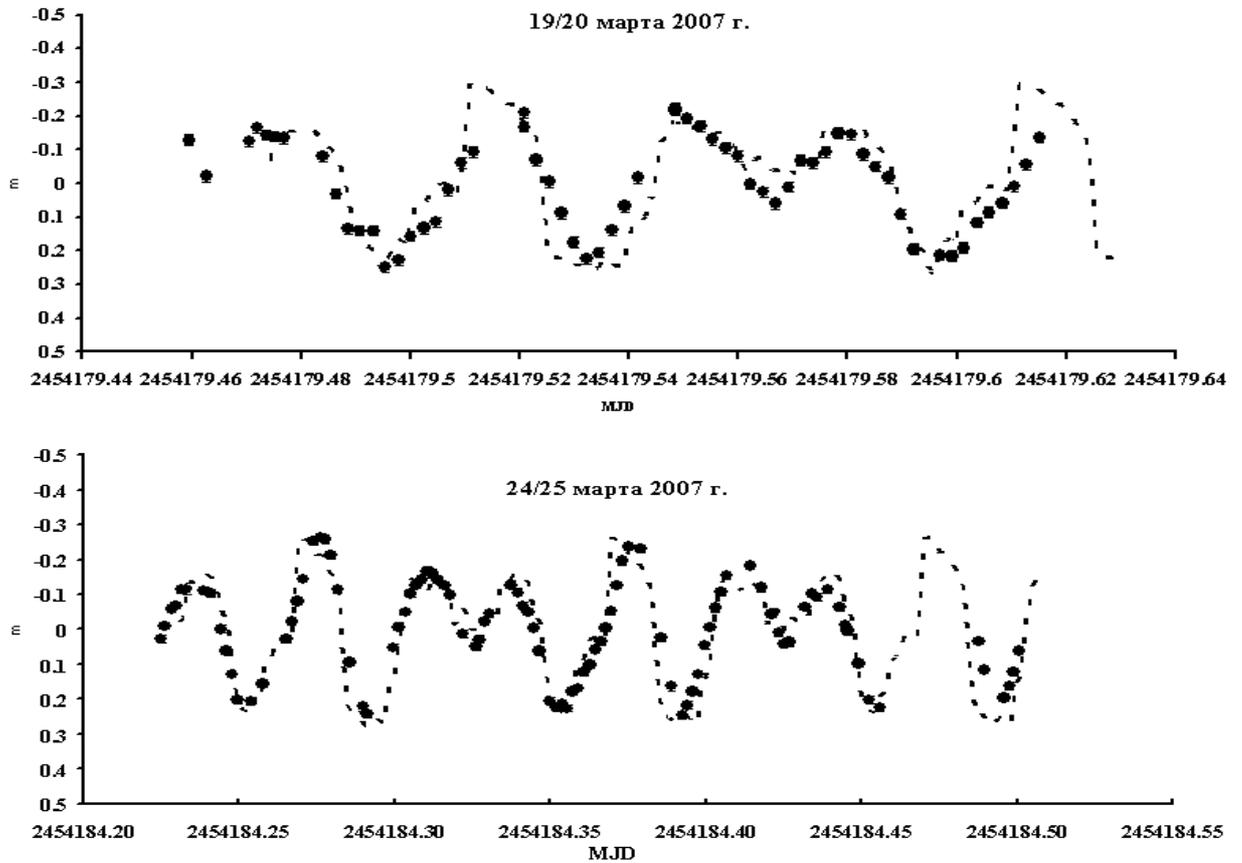

**Fig. 2.** Observed lightcurves of 2006 VV2 asteroid (points) in comparison with the model-curve obtained using the law of the Lumme-Bowell reflection (dotted line).

### 90 Antiope binary asteroid from the main belt

For this asteroid almost sinusoidal light variation with a period of 0.54 years and an amplitude reaching $2^m$ was discovered. This change in brightness is due to the change in phase angle( i.e. area of the illuminated surface of the components) and it deal with the features of the reflective properties of their surfaces [Vereshchagina et al, 2008]. It was simulated the lightcurves of asteroid with the use different laws of reflection. As a result it was defined that the best agreement with the observations gives a lightcurve obtained using the Lumme-Bowell reflection law with the asymmetry factor g = -0.8 (Fig. 3). Also an estimate of slope-parameter for this asteroid was obtained ( G = 0.046 ± 0.023). It is necessary to notice that the Hapke parameters values defined for this asteroid in the paper [Descamps et al, 2009] also are good agreement with observations (Fig. 4).

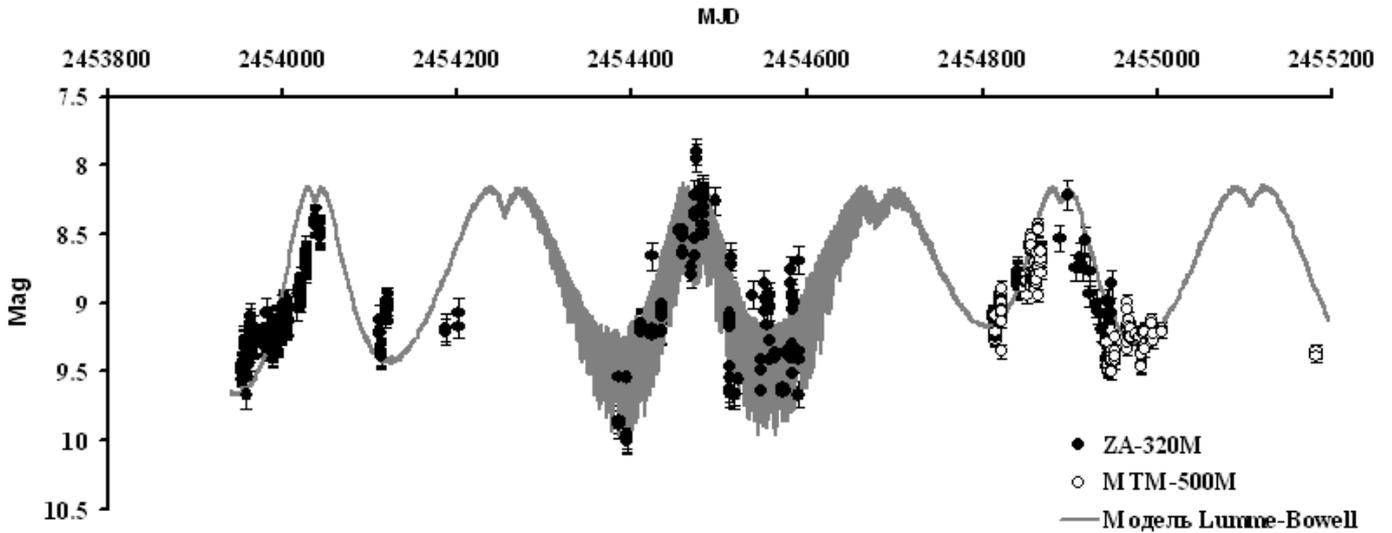

**Fig. 3.** Observations of the 90 Antiope asteroid in the period from 2006 to 2010(points) and model lightcurve obtained using the Lumme-Bowell law of reflection (solid gray line).

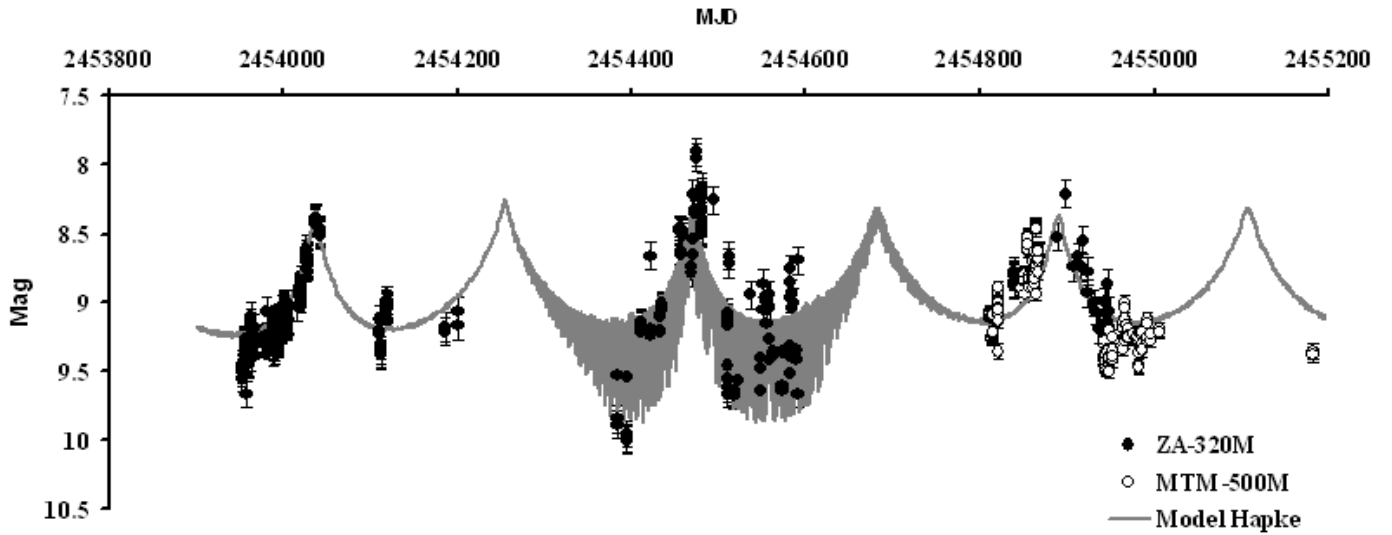

**Fig. 4.** Observations of the 90 Antiope asteroid in the period from 2006 to 2010(points) and model lightcurve obtained using the Hapke law of reflection (solid gray line).

### 45 Eugenia triple asteroid from the main belt

For this asteroid shape of the component has been refined (Figure 5). Direct image of the main component obtained on the Keck II telescope, the previous shape the of main component [Marchis et al, 2006] and a new shape obtained in this work is shown in Figure 6. The coincidence of the model lightcurves for new and old shapes in comparison with the observations is shown in Figure 7. From these results one can see that the shape obtained in this work is in better agreement with observations than the previous one. However, some disagreement with the observational data due to uncertainty in the inclination of the axis of rotation of the main component to the ecliptic remains. This uncertainty with value the 20 degrees reported earlier by other authors.

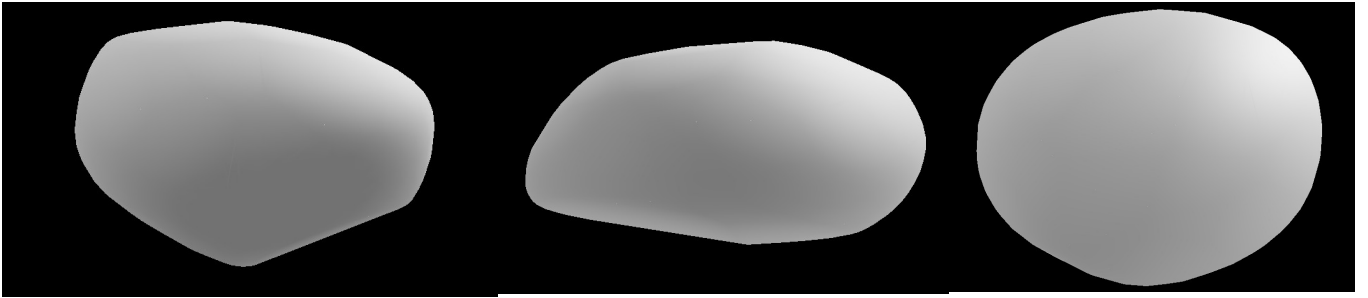

**Fig. 5.** The shape of the main component of the 45 Eugenia asteroid at three different viewpoints, obtained in this work.

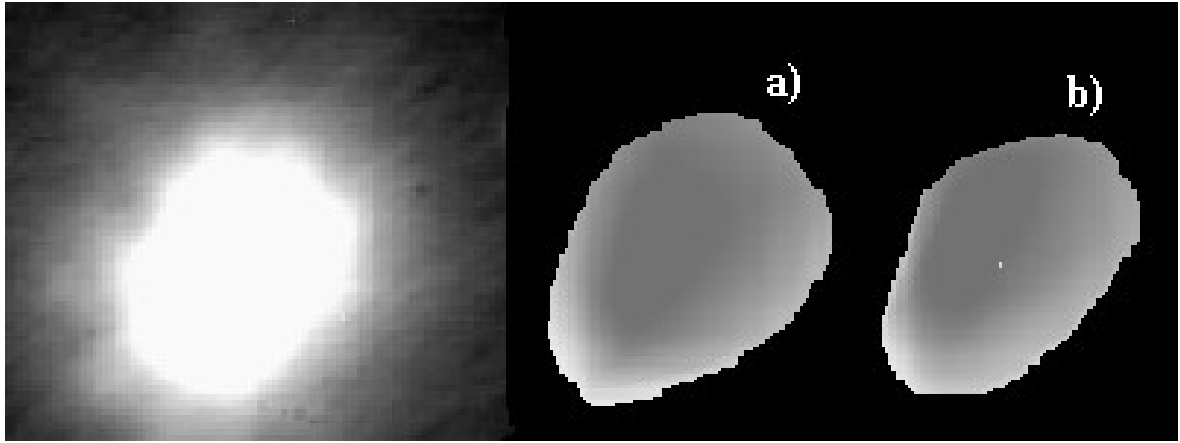

**Fig. 6.** Direct imaging of the main component of the 45 Eugenia asteroid, obtained with telescope with adaptive optics; shape of the main component, obtained in [Marchis et al, 2006] (a) and shap of the main component obtained in this work (b).

On the basis of available data on the asteroid satellites elements of their stable orbits were found . As for the second satellite there were no observational data about its orbit, it has been found possible range of stable orbits, which begins value $a2 = 1.65*a1 = 1930$ km of the semi-major axis, where $a1 = 1170$ km, which is consistent with the results obtained in [Nagy et al, 2010]. The obtained triple system, the evolution of its orbits and the rotation axis of the main component is shown in the Figure 8. It is apparent that the axis of rotation of the main component experiences forced precession associated with the perturbation of the satellites, with an angle of 10 degrees and a period of 66 days. This fact explains the existing uncertainty in the inclination of main component rotation pole to the ecliptic.

Based on a comparison of the model lightcurve with the observations the factor of asymmetry of the asteroid $g = -0.75$ was determined.

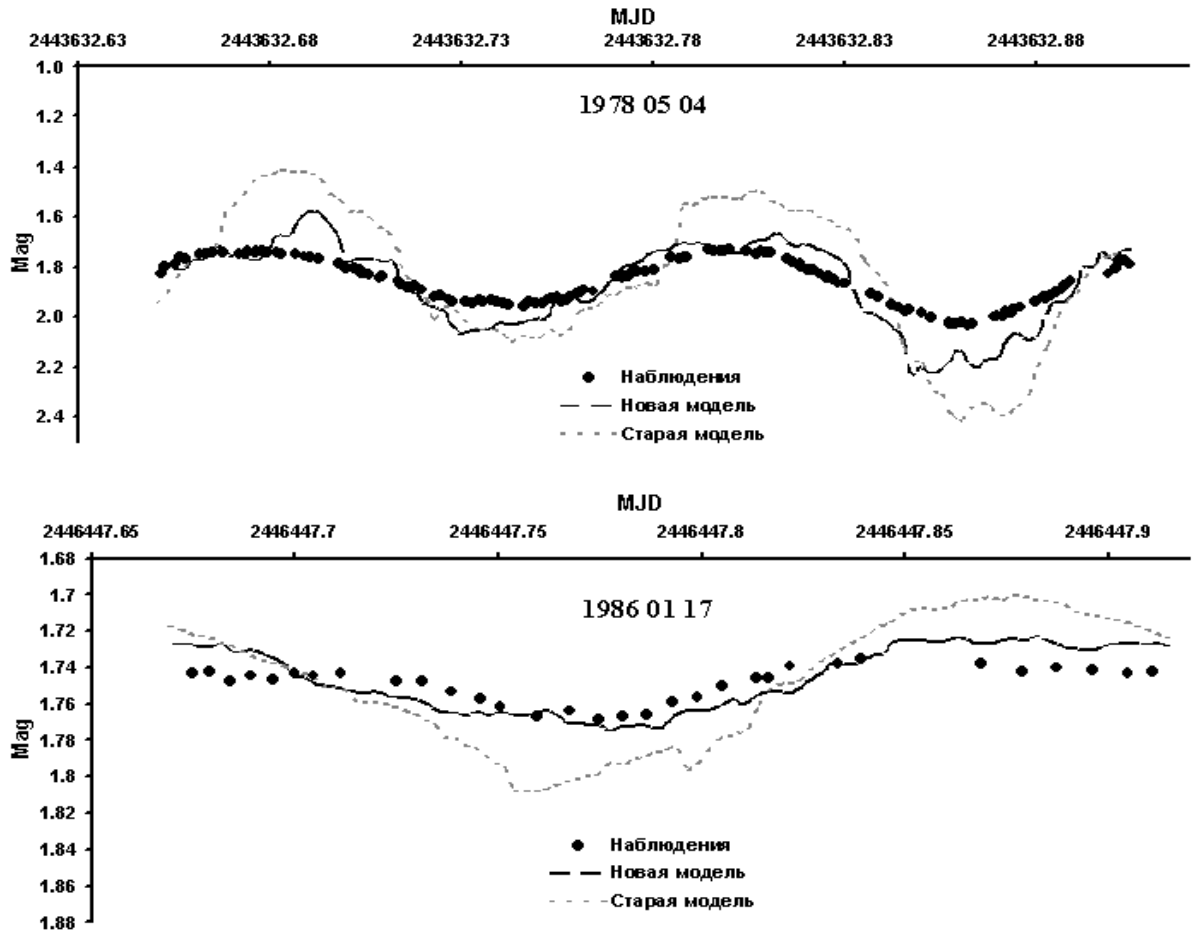

**Fig. 7.** Observations of the asteroid 45 Eugenia (points) compared with model lightcurves obtained for two different shapes of the main components: the previous (gray line) and new (black line).

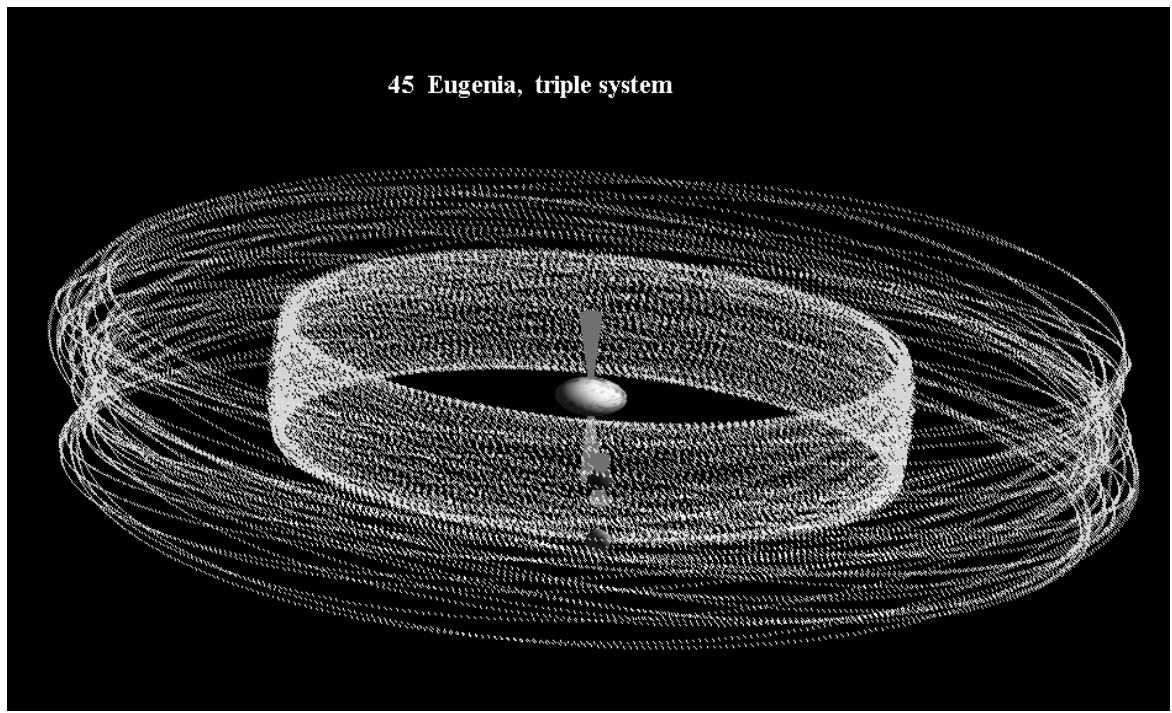

**Fig. 8.** Evolution of the 45 Eugenia asteroid system with two satellites during 1 year.

## Binary main belt asteroid 762 Pulcova

For this asteroid from the observations shape the main component and the pole position of its rotation were identified.. According to the obtained solution, the coordinates of the pole of rotation is β = 71 ± 3 °, λ = 53 ± 2 °.

The resulting shape at three different viewpoints is shown in Figure 9. The comparison of this shape with a direct image of an asteroid obtained with a Keck II telescope [Merlin et al, 2000] is shows in Figure 10 .

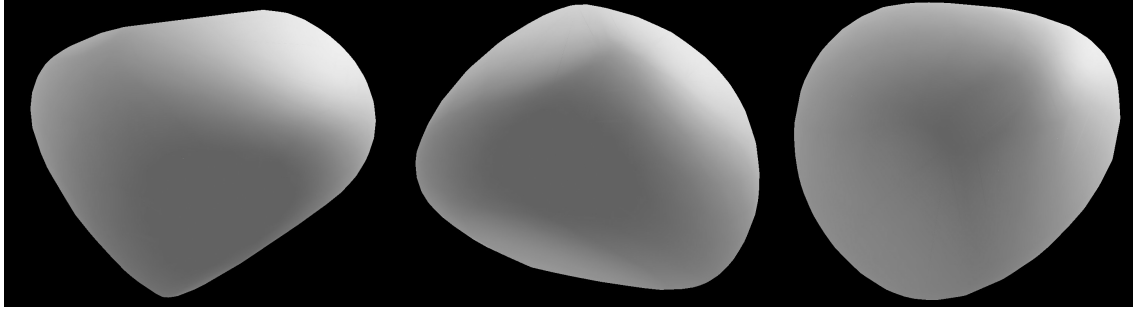

**Fig. 9.** The shape of the main component of 762 Pulcova asteroid, obtained in this work, in three different viewpoints.

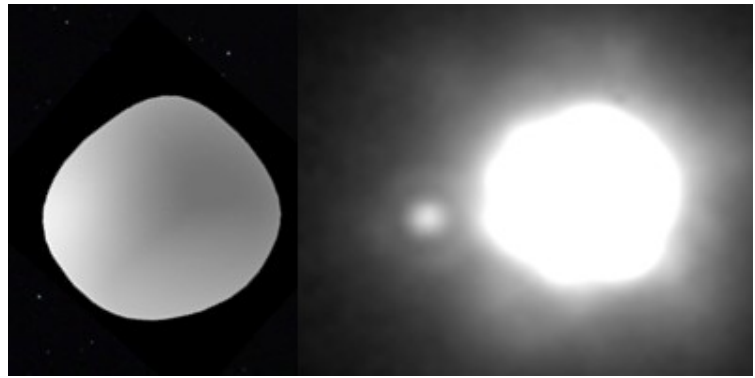

**Fig. 10.** The shape of the main component of 762 Pulcova asteroid , obtained in this work, comp ared with a direct image of an asteroid obtained using a Keck II telescope with adaptive optics .

The model lightcurve obtained using this shape of the main component in comparison with the observations is shown in Figure 11. It is seen that the agreement between model and observations is quite satisfactory.

The results showed that the size of the main component of the asteroid in three dimensions is 74.7 Ч 73.2 Ч 58.9 km, that corresponds to the size ratio $a/b$ = 1.02 and $a/c$ = 1.27.

The elements of a stable orbit corresponding to the observational data were also identified. It was found that elements of the orbit perturb within a € [808, 810] km, e € [0.001, 0.005], i € [0, 2.87] °.

By comparing the model lightcurve with the observations the factor of asymmetry g was determined. The best agreement with observations gives the Lumme-Bowell law of reflection with a value of g = -0.7, as can be seen from Figure 12.

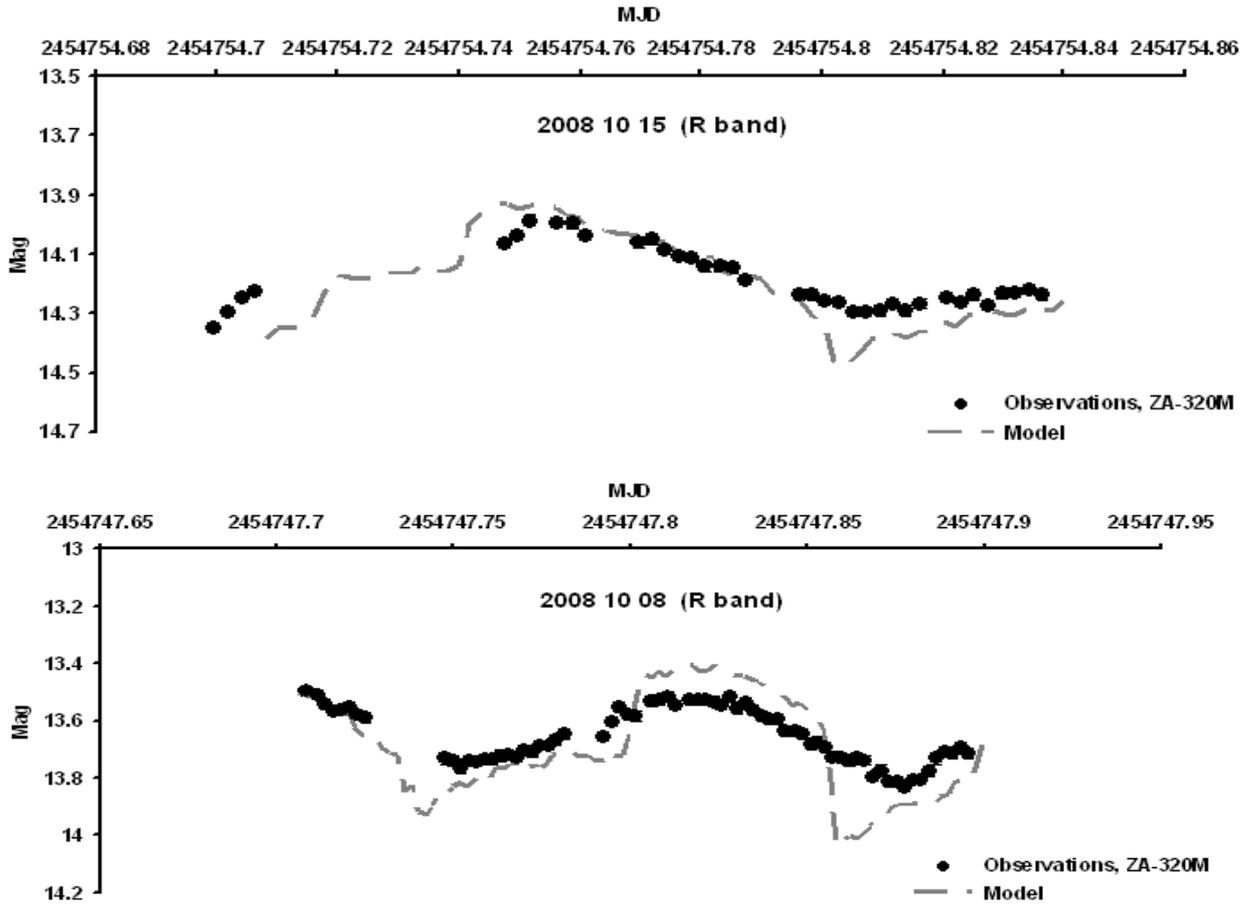

**Fig. 11.** Observations of 762 Pulcova asteroid on different dates compared with model lightcurves.

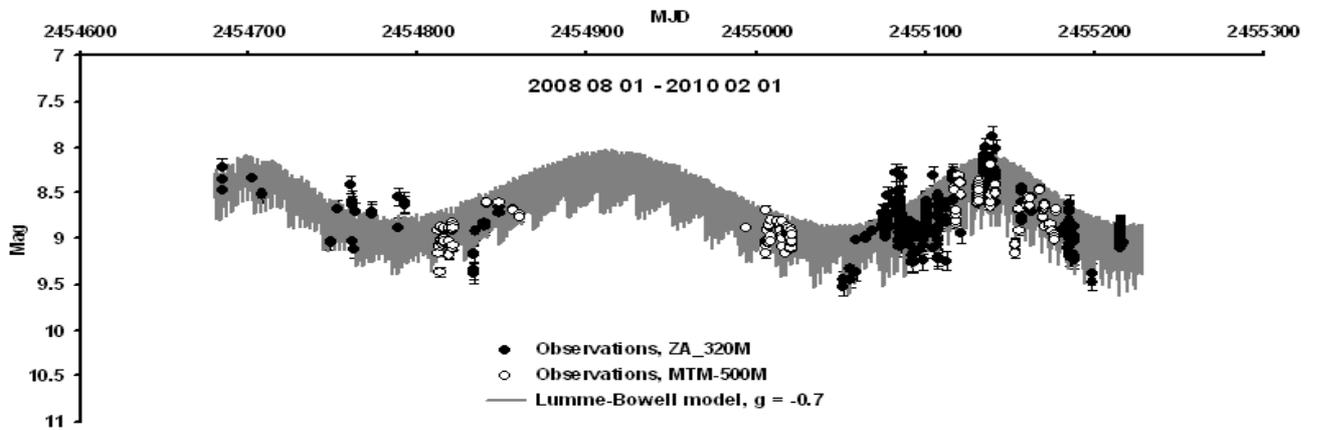

**Fig. 12.** The model lightcurve of 762 Pulcova asteroid for Lumme-Bowell reflection law (gray line) compared with observations (points).

### 137170 (1999 HF1) binary asteroid from NEA group

Estimates of main component size were obtained using the observation data: $a = 2.12 \pm 0.02$, $b = 1.77 \pm 0.01$, $c = 1.73 \pm 0.03$ km. Possible stable orbit of the satellite with the elements of $a \sim 6.2$ km, $e \sim 0.1$, $i \sim 2°$, which corresponds to the observational data most closely, have been defined. Using these results and Kepler's third law, were also made estimates of the components, masses which are given in Table 6.

**Table 6.** Estimates of the components masses for 137170 (1999 HF1) asteroid.

|  | Mass, kg |
|---|---|
| **Main component** | $5.43 \times 10^{13} \pm 0.12 \times 10^{13}$ |
| **Satellite** | $8.15 \times 10^{11} \pm 0.05 \times 10^{11}$ |
| **System mass** | $5.5 \times 10^{13} \pm 0.1 \times 10^{13}$ |

Based on the comparison of observations with the model lightcurve was established that the value of asymmetry factor for a given asteroid is very large in absolute value: $g = -0.9$.

### Triple main belt asteroid 87 Sylvia

For this asteroid using the large observational material obtained in this work, the shape of the main component has been refined. This obtained shape was significantly different from previous one estimate [Marchis et al, 2006]. The both shapes are shown in comparison with direct images of asteroid in Figure 13. It is clear that the new shape gives better agreement with observations. The same is confirmed by Figure 14, which provides a comparison of model lightcurves for different shapes witn the observations.

The comparison of model lightcurves for different shapes with observations during a longer time interval, as is evident from Figure 15, also shows that the new shape gives better agreement with observations data. However, both shapes are marked discrepancies with the observations. This suggests that a more precise study of the asteroid using additional observational data is needed.

From the lightcurves shown in Figure 16, it was found that the best match with observations is obtained by using the Lumme-Bowell law of reflection with a value of asymmetry factor $g = -0.7$.

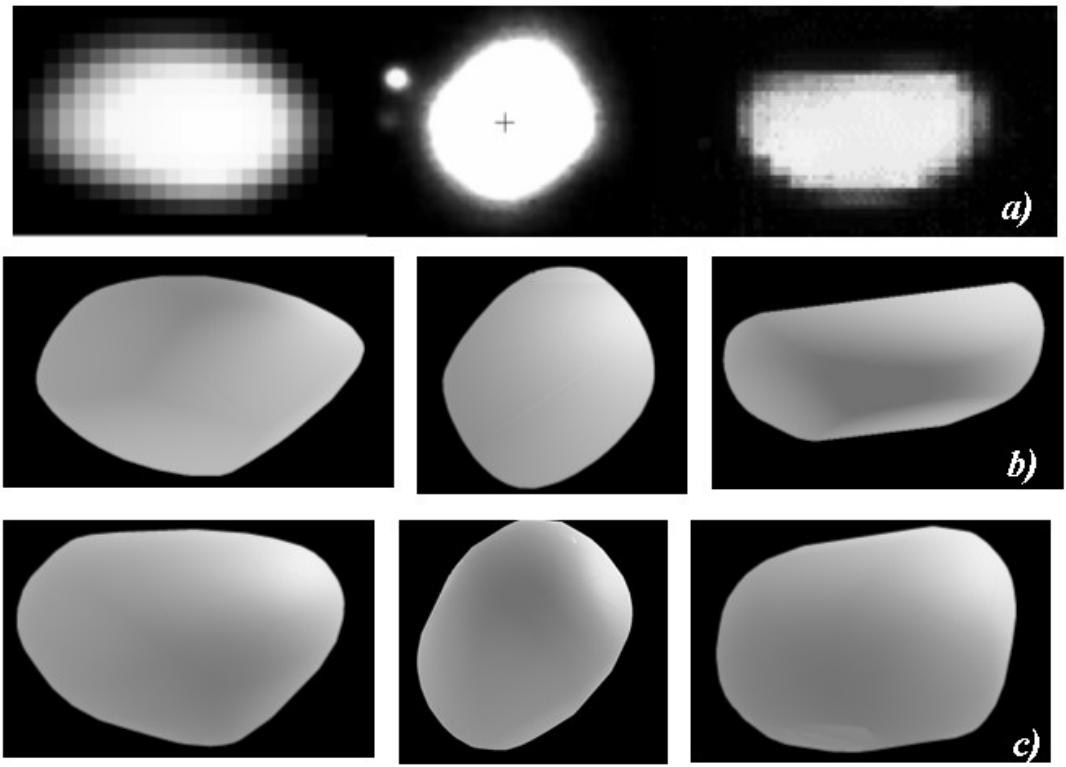

**Fig. 13.** Direct images of 87 Sylvia asteroid, obtained by Keck-II telescope (a), a new shape of main component obtained in this work (b), the previous shape the main component (c).

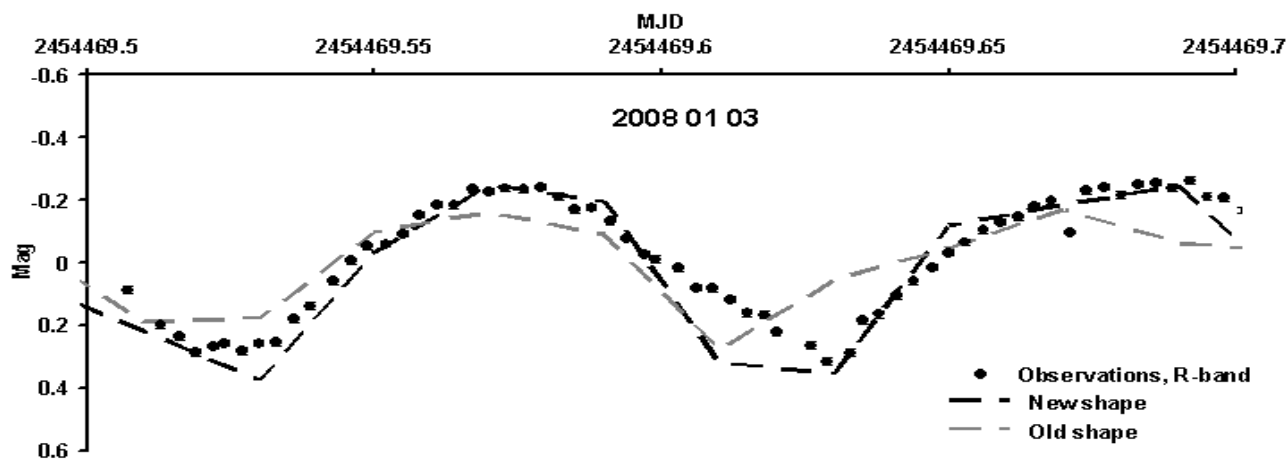

**Fig. 14.** Observed lightcurve of 87 Sylvia asteroid (points) compared with model lightcurves obtained using two different shapes of the main components: the previous (gray line) and new (black line).

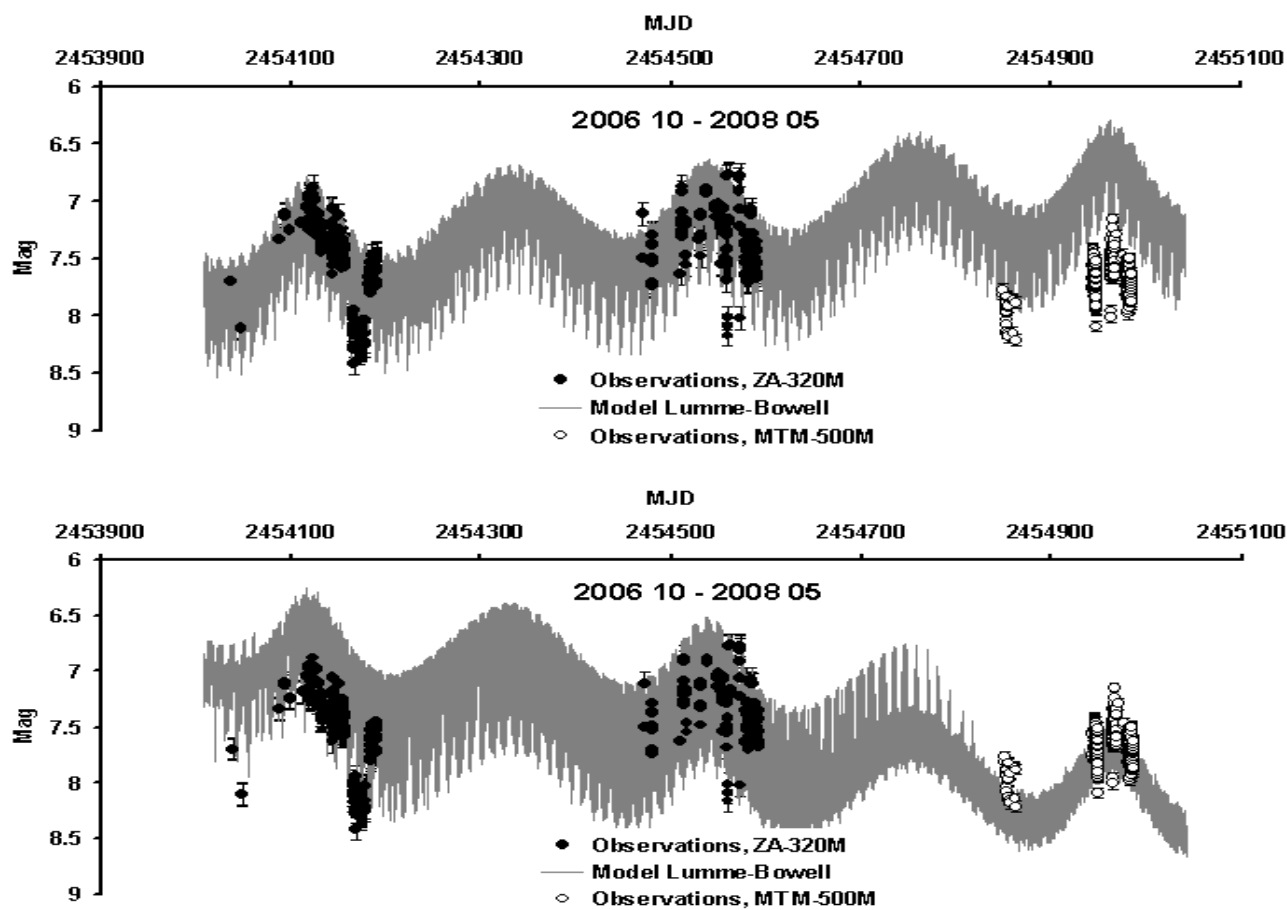

**Fig. 15.** Model lightcurves for the previous (top) and new (lower) shapes the main component(gray line) compared with observations (points).